\documentclass[]{spie} 

\usepackage[]{graphicx}
\usepackage{xspace} 				
\usepackage{gensymb}        
\usepackage[]{enumitem} 

\usepackage{color}
\definecolor{blue}{rgb}{0.1,0.1,0.6}
\definecolor{orange}{rgb}{0.74,.35,0.099}
\definecolor{pale}{rgb}{0.90,0.90,0.95}
\definecolor{red}{rgb}{1.0,0.0,0.0}

\newcommand{\um}[0]{$\mu$m\xspace}

\title{James Webb Space Telescope Optical Simulation Testbed I: Overview and First Results}

\author{Marshall D. Perrin\supit{a}, 
{R\'emi Soummer}\supit{a},
\'Elodie Choquet\supit{a},
Mamadou N'Diaye\supit{a},\\
Olivier Levecq\supit{a,b},
Charles-Phillipe Lajoie\supit{a},
Marie Ygouf\supit{a},
Lucie Leboulleux\supit{a,b},
Sylvain Egron\supit{a,b},
Rachel Anderson\supit{a},
Chris Long\supit{a},
Erin Elliott\supit{a},
George Hartig\supit{a},
Laurent Pueyo\supit{a},
Roeland van der Marel\supit{a},
Matt Mountain\supit{a}
\skiplinehalf
\supit{a} Space Telescope Science Institute, 3700 San Martin Dr, Baltimore, MD 21218 USA; \\
\supit{b} Institute d'Optique Graduate School, Palaiseau, Saint-\'Etienne and Bordeaux, France
}

\authorinfo{Please send correspondence to M.D.P. at mperrin@stsci.edu.}

\begin{document} 
 \maketitle 

\begin{abstract}
The James Webb Space Telescope (JWST) Optical Simulation Testbed (JOST) is a tabletop workbench to study aspects of wavefront sensing and control for a segmented space telescope, including both commissioning and maintenance activities. JOST is complementary to existing optomechanical testbeds for JWST (e.g. the Ball Aerospace Testbed Telescope, TBT) given its compact scale and flexibility, ease of use, and colocation at the JWST Science \& Operations Center. We have developed an optical design that reproduces the physics of JWST's three-mirror anastigmat using three aspheric lenses; it provides similar image quality as JWST (80\% Strehl ratio) over a field equivalent to a NIRCam module, but at HeNe wavelength. A segmented deformable mirror stands in for the segmented primary mirror and allows control of the 18 segments in piston, tip, and tilt, while the secondary can be controlled in tip, tilt and x, y, z position. This will be sufficient to model many commissioning activities, to investigate field dependence and multiple field point sensing \& control, to evaluate alternate sensing algorithms, and develop contingency plans. Testbed data will also be usable for cross-checking of the WFS\&C Software Subsystem, and for staff training and development during JWST's five- to ten-year mission. 

\end{abstract}



\section{INTRODUCTION}
\label{sec:intro} 

Active optical control based on wavefront sensing is an essential technology that has revolutionized terrestrial telescopes and enabled the current generation of 8--10 meter observatories. While the Hubble Space Telescope has a classical, mostly passive optical design with only occasional adjustments for focus, its successor the James Webb Space Telescope (JWST) would not be possible without active control of its primary and secondary mirrors.  Active optical control enables bringing the segmented primary from an initial deployed state with millimeter misalignments to a final aligned state precise to tens of nanometers. Subsequent periodic wavefront sensing and control is needed to maintain that aligned state against perturbations for the lifetime of the mission 
\cite{Acton2004SPIE.5487..887A,Acton2012SPIE_JWSTWFSC_Overview,Knight2012SPIE.8442E..2CK}. 
Wavefront sensing and control (WFS\&C) technologies are likewise critical ingredients in future large space telescopes such as the proposed WFIRST/AFTA\cite{Spergel2013arXiv1305.5422S} and ATLAST\cite{Postman2009arXiv0904.0941P} mission concepts. 

The wavefront sensing and control technologies chosen for JWST have been proven by a long-term technology development plan of simulations and laboratory demonstrations\cite{Barto2008SPIE.7010E..23B}, in particular using the Testbed Telescope (TBT) at Ball Aerospace\cite{Acton2006SPIE.6265E..21A,Acton2007SPIE.6687E...5A}, a 1:6 scale mockup of JWST with flight-like segmentation and actuation, and the Integrated Telescope Model (ITM) software \cite{Knight2012SPIE.8449E..0VK}. However, new and improved algorithms for wavefront sensing and control continue to be developed 
\cite{Sivaramakrishnan2012SPIE.8442E..2SS,Jurling2012SPIE.8442E..10J,Pope2014MNRAS.440..125P,Jurling:2014}, and a flexible general-purpose lab testbed will contribute to evaluating such algorithms for possible application to JWST or to future missions such as WFIRST/AFTA. Furthermore, there is a continued need to practice scenarios and develop staff expertise for JWST commissioning and operations, including in the longer run a need for occasionally training new staff in phase retrieval techniques throughout the five-to-ten year mission of JWST. 


For these reasons we have developed a new laboratory testbed for wavefront sensing and control at STScI, with emphasis on JWST but intended more broadly as a testing ground for novel wavefront control methods with application to future space missions. The testbed design provides a simplified model of JWST which is sufficient to capture the key optical physics of JWST (a three-mirror anastigmat design and segmented primary), and to implement sensing and control algorithms applicable to flight-like scenarios. Our goal is not to replicate the higher fidelity model offered by the TBT, but instead to provide a compact, flexible, low cost ``tabletop JWST'' that nonetheless possesses enough of the same physics to be a useful model of controlling an active segmented space telescope. This testbed, which we have termed the JWST Optical Simulation Testbed (JOST) has been developed by STScI's Telescopes Team and is physically located within the Russell B. Makidon Optics Laboratory at STScI. It is a supplement to existing verification and validation activities for independent cross-checks and novel experiments, not a part of the mission's critical path development process.

This paper presents the overall motivation, goals, and requirements for this testbed, and some first initial results. For the detailed optical design and trade studies, see the accompanying paper by Choquet et al.~in these proceedings\cite{Choquet2014}, paper number 9143-143 (hereafter denoted ``Paper II"). JOST was designed and developed starting in mid-2013, with a prototype using off-the-shelf lenses assembled by September 2013, and the final design with custom lenses assembled by end of 2013. Further alignment and test work is ongoing.

Planned activities using this testbed include: 
\begin{itemize}\setlength{\itemsep}{1pt}
\setlength{\itemsep}{1pt}
\item Simulation of the main commissioning and maintenance tasks for JWST WFS\&C, including both single field point and multi-point control analogous to the Multi-Instrument Multi-Field\cite{Acton2012SPIE_MIMF} control step for JWST.
\item Experimentation with alternate algorithms for wavefront sensing, either as enhancements providing superior performance or capture range compared to baseline algorithms\cite{Jurling:2014} or as contingency plans for backup\cite{Sivaramakrishnan2012SPIE.8442E..2SS}.
\item Hands-on training to develop additional staff expertise in WFS\&C, both prior to launch and as a resource for training new staff during the mission lifetime of JWST.
\item Potentially using the testbed as a resource to exercise the integrated Science \& Operations Center ground system and Wavefront Sensing Subsystem (WSS) software prior to launch, to supplement the planned tests of the WSS during I\&T using the flight hardware and separate planned tests using simulations from the Integrated Telescope Model\cite{Knight2012SPIE.8449E..0VK}.
\end{itemize}

\section{TESTBED DESIGN OVERVIEW} 

\subsection{Design Process and Goals}
Even though JOST is a multipurpose lab testbed rather than any sort of mission-critical flight hardware, we approached its design as an exercise in science systems engineering following the same general principles as commonly used for flight systems. JOST was developed by a small team of staff astronomers, postdocs, and graduate students. We first developed a clear set of requirements, iterated through multiple designs and conducted trade studies exploring parameter space, and conducted a mini design review. 

We began the detailed optical design of JOST based on the following top-level goals:
\begin{itemize}[itemsep=3pt]
\item The testbed must be capable of a wide variety of wavefront sensing and control experiments relevant to JWST, with an optical design that replicates the relevant physics and aspects of the JWST optical design.
\item In particular it must be possible to carry out both fine phasing at a single field point, and control based on multiple field points in the manner of the Multi-Instrument Multi-Field (MIMF) step of JWST commissioning.
\item It must make use of an engineering-grade micro­electro­mechanical system (MEMS) segmented deformable mirror (DM) from Iris AO\footnote{http://www.irisao.com}, which we already had available in the lab from the HICAT\cite{2013SPIE.8864E..1KN} project. 
\end{itemize}
We then defined a set of 10 requirements necessary to achieve these goals (see Paper II for more details). 
\noindent 
After some consideration of alternatives, we chose to pursue a transmissive design that approximates a three mirror anastigmat (TMA) system very similar to the actual JWST layout. The use of refractive optics allowed significant cost savings compared to implementing a reflective TMA given that there is no need to operate over a broad spectral bandpass. The starting point for the design was thus the JWST design\cite{Knight2012SPIE.8442E..2CK} for M1, M2, M3.  The re-imaged pupil where JWST's Fast Steering Mirror (FSM) is located provides a convenient pupil at which to place the MEMS DM.  Paper II discusses in detail the solution to this optical design problem.

\subsection{General Implementation}
\begin{figure}[t]
 \centering
  \includegraphics[height=5cm]{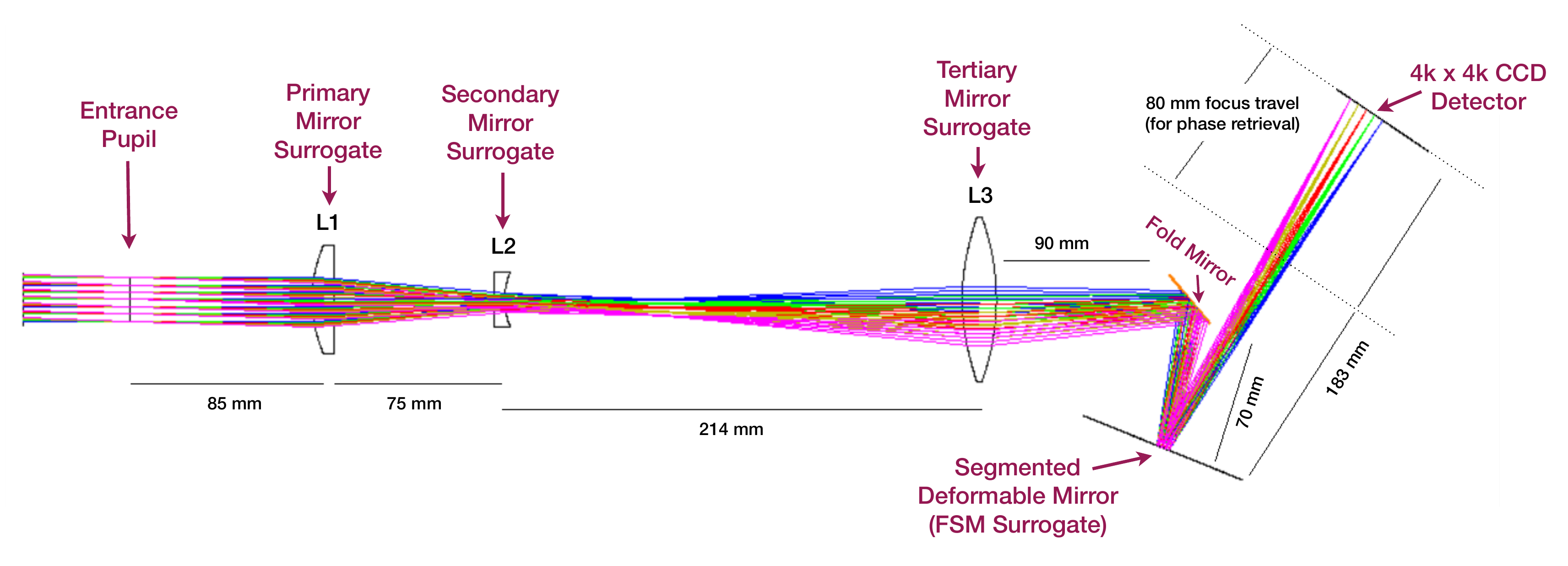}
\caption{\label{fig:design}Optical layout for JOST, with components labeled. A collimated beam enters the system from left, passes through the three refractive optics and then reflects off the segmented deformable mirror to the detector. A flat fold mirror is included before the deformable mirror for packaging reasons to provide sufficient clearance between the detector and the L3 lens. The detector can be translated up to 8 cm to obtain defocus for phase retrieval.}
\end{figure}

The final optical design is a ``three lens anastigmat'' analogous to a refractive version of JWST's three-mirror anastigmat optical design.  See Figure \ref{fig:design}. Controllable modes are set by the segmented Iris AO DM and the secondary mirror surrogate. Translation of the CCD camera is used to implement focus-diverse phase retrieval, and a Fizeau interferometer (4D AccuFiz) can be used as a direct wavefront sensor to validate our implementation of wavefront sensing algorithms. The testbed is designed to work at visible light (HeNe laser wavelength). 

The JWST primary mirror surrogate consists of three elements that together mimic the real telescope:
\begin{itemize}[itemsep=2pt]
\item An Iris AO 37 ­segment MEMS DM with 1.4~mm (vertex to vertex) hexagonal segments controllable in tip, tilt and piston, conjugate to the entrance pupil. Note that this size of device is a new product from Iris AO, scaled up $2\times$ to 1.4 mm per segment instead of the previous 0.7~mm. The larger device provides significant relaxation of pupil alignment tolerances compared to the smaller mirror.
\item An entrance pupil mask to limit the beam to 18 active segments of the 37 total in the DM, and mask out the telescope central obstruction and support structures. The use of an engineering-grade DM with one or more non-functional segments would in fact be sufficient, provided that 18 suitably adjacent segments are functional.\footnote{We are currently using an engineering grade mirror with 17 functional segments in the JWST aperture (nonfunctional center segment and one adjacent).  This is sufficient for the majority of currently planned activities for the immediate future. We expect to take delivery of two science-grades device later this year with all segments operational. } 
\item A lens to provide the overall optical power for the primary, since the Iris AO segments are flat. 
 \end{itemize}

The entrance pupil corresponds to a circumscribed diameter of 6.1~mm on the DM.
The segment gap size is 10--­12~\um on the MEMS DM ($\sim0.1\%$ of segment diameter) which makes the gaps to scale relative to the actual JWST geometry. The surface quality of each segment is slightly better than JWST ($\sim20$~nm rms surface error, versus 23.2 nm rms measured for the 18 flight PMSAs for JWST) but the testbed will operate at shorter wavelength in the visible, so that this setup will in the end provide a reasonable model of the telescope compatible with our goals.

The JWST secondary mirror surrogate consists of a diverging lens. As with the actual telescope it ``slows down'' the beam and produces an intermediate focal plane. The lens is mounted on a motorized tip/tilt and translation stage to study the alignment of the telescope. 

The JWST tertiary mirror surrogate consists of a converging aspherical lens. It creates the final image on the detector with the appropriate f-­ratio calculated to produce the same sampling as NIRCam relative to the diffraction-limited PSF size at the wavelength of interest. 

The controllable degrees of freedom (DOF) of the testbed include tip, tilt, and piston for the primary mirror (54 DOF), and tip, tilt, $x$, $y$, and $z$ for the secondary mirror (5 additional DOF; 59 total). This does not reproduce all the degrees of freedom of the actual telescope (131 DOF) since the segments' radius of curvature, translations, and clocking are not included. However the 54 DOF are sufficient to implement a number of commissioning activities (e.g. segment identification, image array, global alignment, stacking, fine phasing) and to investigate wide-field WFS\&C. 

The selected detector is a 4k~$\times$~4k SBIG STX-­16803 CCD with $9~\mu$ pixel (36~mm length per side of the focal plane), providing the same number of pixels as a NIRCam short-wavelenth channel.\footnote{Admittedly this setup does not replicate the small gaps between detectors in the $2\times2$ mosaic of Hawaii 2RGs in NIRCam, but that is a nonessential detail with no direct impact on wavefront sensing.}  To reproduce a similar field of view as one NIRCam module in the image plane measured in units of $\lambda/D$, this testbed has an actual field of view of $3.4\degree \times 3.4\degree$, which is significantly greater than the field of view of the NIRCam detectors ($2.2' \times 2.2'$). Based on our error budget allocation for the wavefront error with a goal of 50~nm rms total WFE for diffraction-limited imaging at HeNe, and including contributions from the lens surfaces, Iris AO DM, and mechanical alignments, the allocated WFE for the design itself over the field of view is 10--20~nm rms. Achieving this very small WFE over such a wide field was a very challenging and interesting optical design problem, one however that was solved successfully using three custom lenses, with a 4th order aspherical term on each. See Paper II for details on this design.

We briefly considered operating at longer wavelengths ($\sim$ 750~nm) to relax WFE tolerances. However in order to have the same sampling as NIRCam, working at 750~nm would require a faster lens for L3 (f-ratio = 25.4, instead of 30.2 at HeNe). We found that this would be problematic for achieving sufficient clearance between the DM and the camera. Furthermore the shorter wavelength offers better sensitivity both for the human eye and for our chosen detector's quantum efficiency.

For illuminating the entrance pupil, two flat mirrors (one beam capture mirror, and one beam steering mirror) allow steering an oversized collimated beam to different angles in order to simulate stars at different positions within the field of view. The steering mirror is motorized to facilitate the implementation of multi-field WFS\&C studies. The input collimated beam may be provided either using the 4D AccuFiz interferometer or a laser launched from a fiber and collimated by an off-axis parabola (OAP). 

\section{ASSEMBLY, STATUS, AND INITIAL RESULTS}

\subsection{Hardware Status And Assembly}

\begin{figure}[t]
 \centering
  \includegraphics[width=0.8\textwidth]{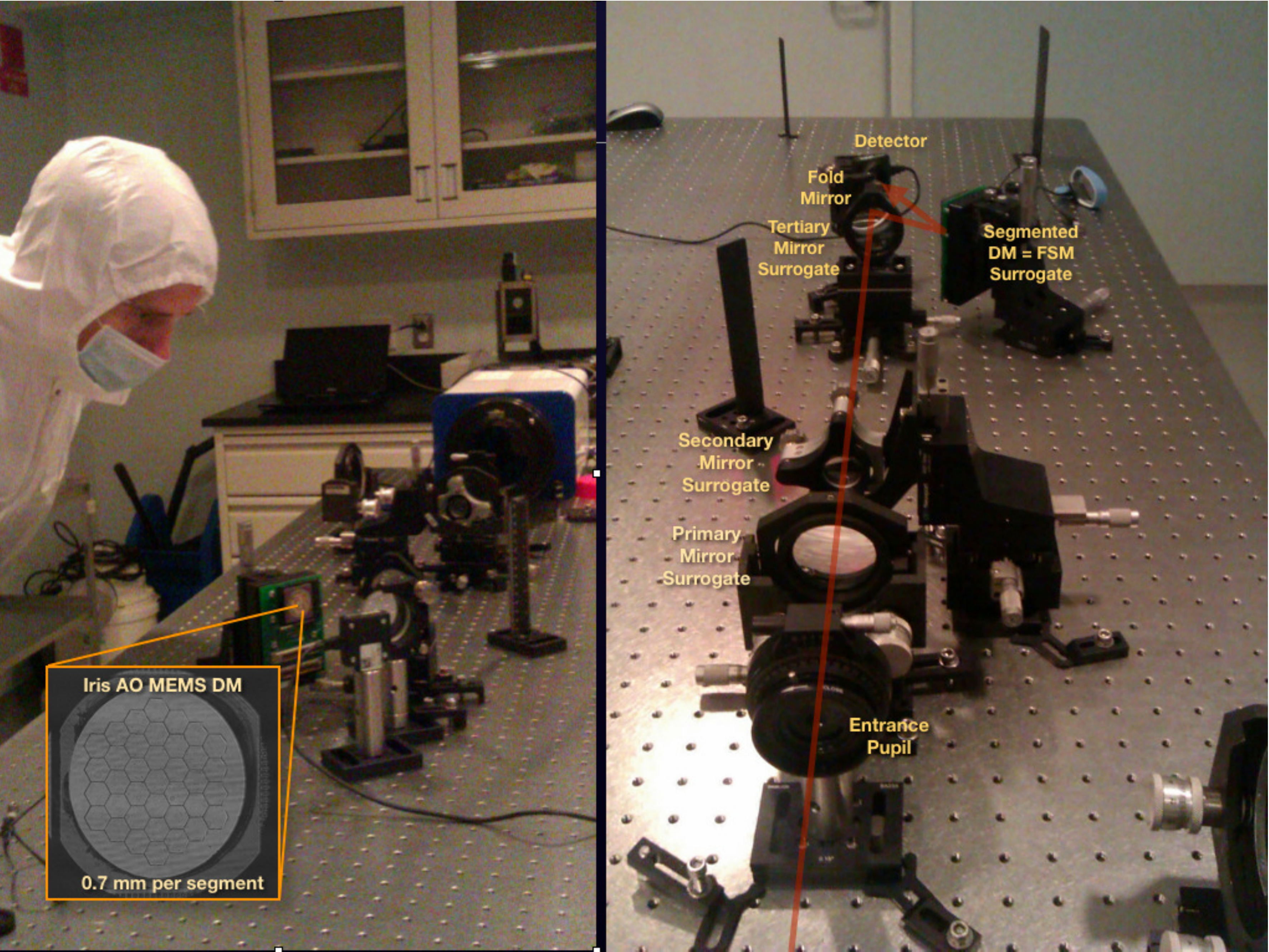}
   \caption{\label{fig:hardware_prelim}  Photos of the preliminary version testbed with off-the-shelf lenses for the optimization of the mechanical design and preliminary alignment, as seen from opposite ends of the system. \textit{Left:} Looking back from above the location of the detector (not yet installed when this photo was taken). A zoomed in view of the deformable mirror is inset lower left; the preliminary testbed used the stock 0.7 mm per segment Iris AO DM, as opposed to the newer 1.4 mm per segment devices selected for the final testbed. The 4D AccuFiz interferometer is visible in the distance (the white and blue rectangular object). \textit{Right}: Looking in the opposite direction from above the entrance aperture. The light path and optics are annotated. In both of these images a small video-rate alignment camera is installed in place of the 4k CCD.}
\end{figure}

\begin{figure}[t]
 \centering
  \includegraphics[width=0.9\textwidth]{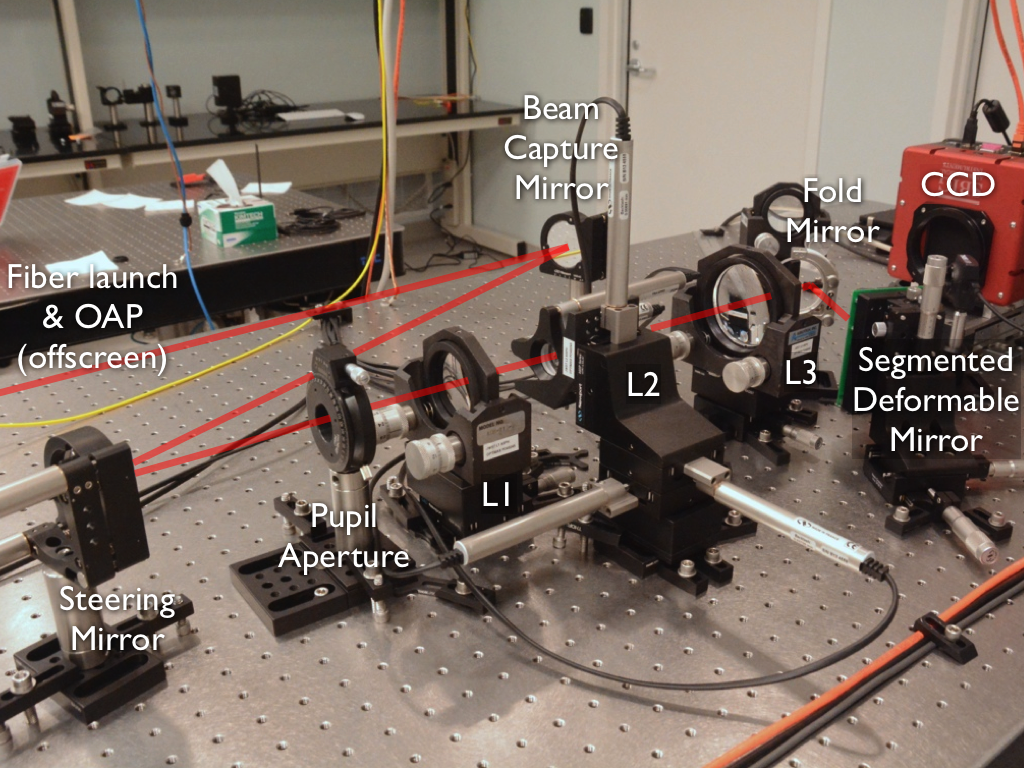}
   \caption{\label{fig:hardware_final} The final JOST testbed assembled including all components. The red line indicates the beam path. Laser light collimated from an OAP enters from left and is steered by a pair of flats into the entrance pupil; the same beam capture mirror can be used to inject the 4D interferometer beam for direct wavefront sensing. The L1, L2, and L3 custom aspheric lenses can be seen in their mounts. The L2 mount is motorized on 5 axes for active control of the secondary, and the steering mirror is motorized in tip and tilt for moving the ``star'' around the field of view.  At right a fold mirror directs the light onto the segmented DM (facing away from us) which reflects it toward the CCD for detection. The additional optic seen behind the fold mirror, in the black mount, is an pupil imaging lens that can be manually inserted into the beam when needed. }
\end{figure}

Based on the final optical design, we developed and optimized the optomechanical design in Solidworks CAD software\footnote{http://www.solidworks.com} using off-the-shelf mounts and stages, including 5-axis motorization of the secondary mirror surrogate (tip, tilt, $x, y, z$),  2-axis motorization (tip, tilt) of the beam steering mirror to facilitate multi-field imaging, and translation of the camera for focusing and phase retrieval.

Because of the long lead time associated with the custom aspheric lenses needed for the final design, a simplified ``preliminary'' version of the optical design was developed using off-the-shelf lenses to achieve our requirements on-axis (though with very substantial off axis aberrations). This was assembled starting in August 2013 in order to provide a prototype for initial tests with the deformable mirror and validation of the optomechanical parts, and is shown in Figure \ref{fig:hardware_prelim}.

The starlight simulator consists of a fiber optic point source at the focal point of a custom high-quality off-axis parabola (NuTek Precision Optical Corp., measured to 13 nm rms = $\lambda/40$ at 633 nm over the 2 inch clear aperture), precisely aligned with the help of the 4D interferometer.  We adopted a design that uses two high quality flat mirrors ($\lambda/20$ peak-to-valley over a 2 inch clear aperture, measured with the 4D to be just a few nm rms each over the 18 mm entrance pupil) to capture and steer the beam into the entrance aperture. The high quality optics ensure the starlight simulator is a negligible contributor to the overall wavefront error budget. This beam capture relay can also conveniently redirect the 4D interferometer's collimated beam for interferometric measurements; there is sufficient space on the table to simply place the 4D in front of the fiber launch and collimator without needing to move those optics or disturb their alignment in any way.  As noted above the beam steering mirror is motorized in tip and tilt to allow steering the ``star'' to any location within the field of view.

\begin{figure}[t]
 \centering
  \includegraphics[height=2.5in]{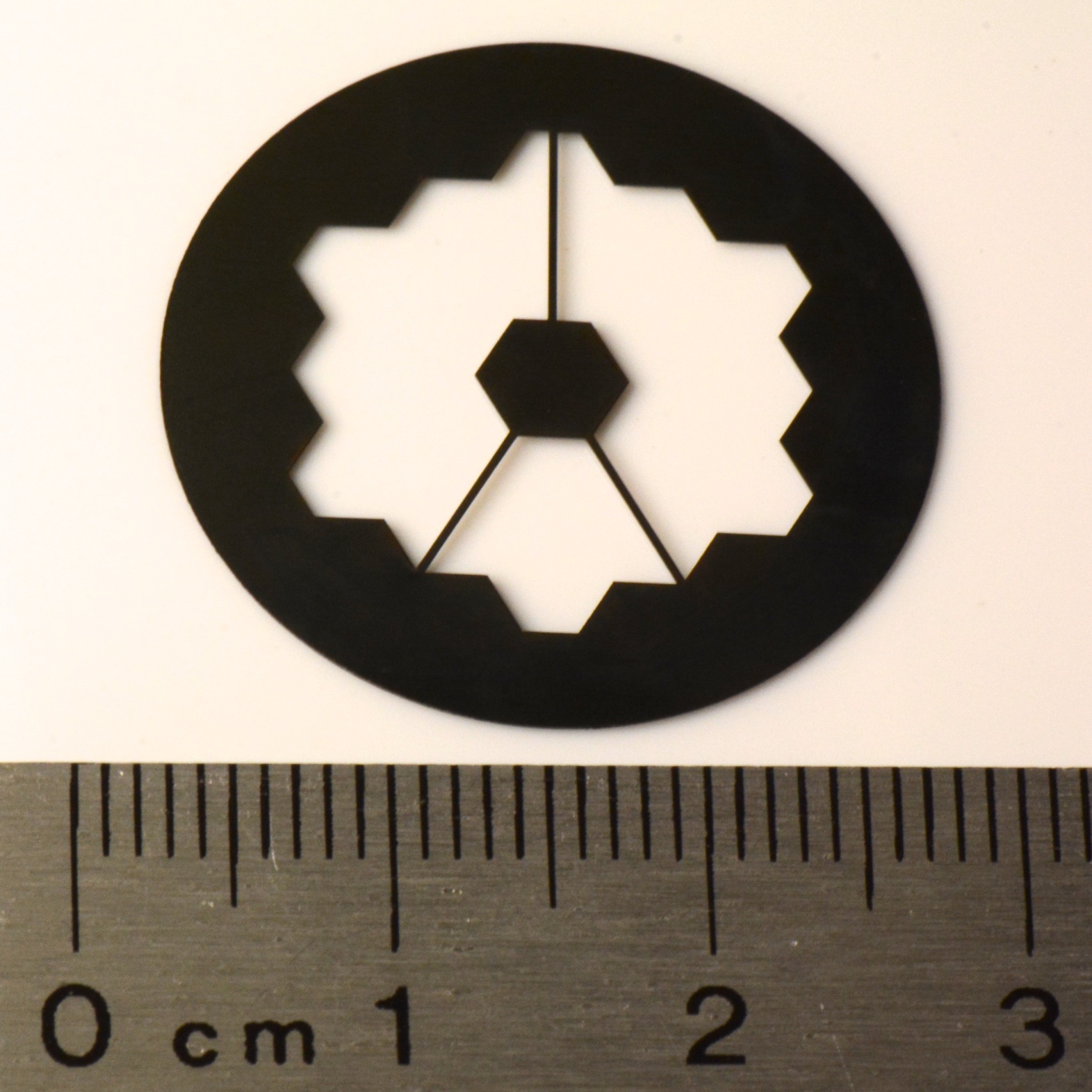}
  \includegraphics[height=2.5in]{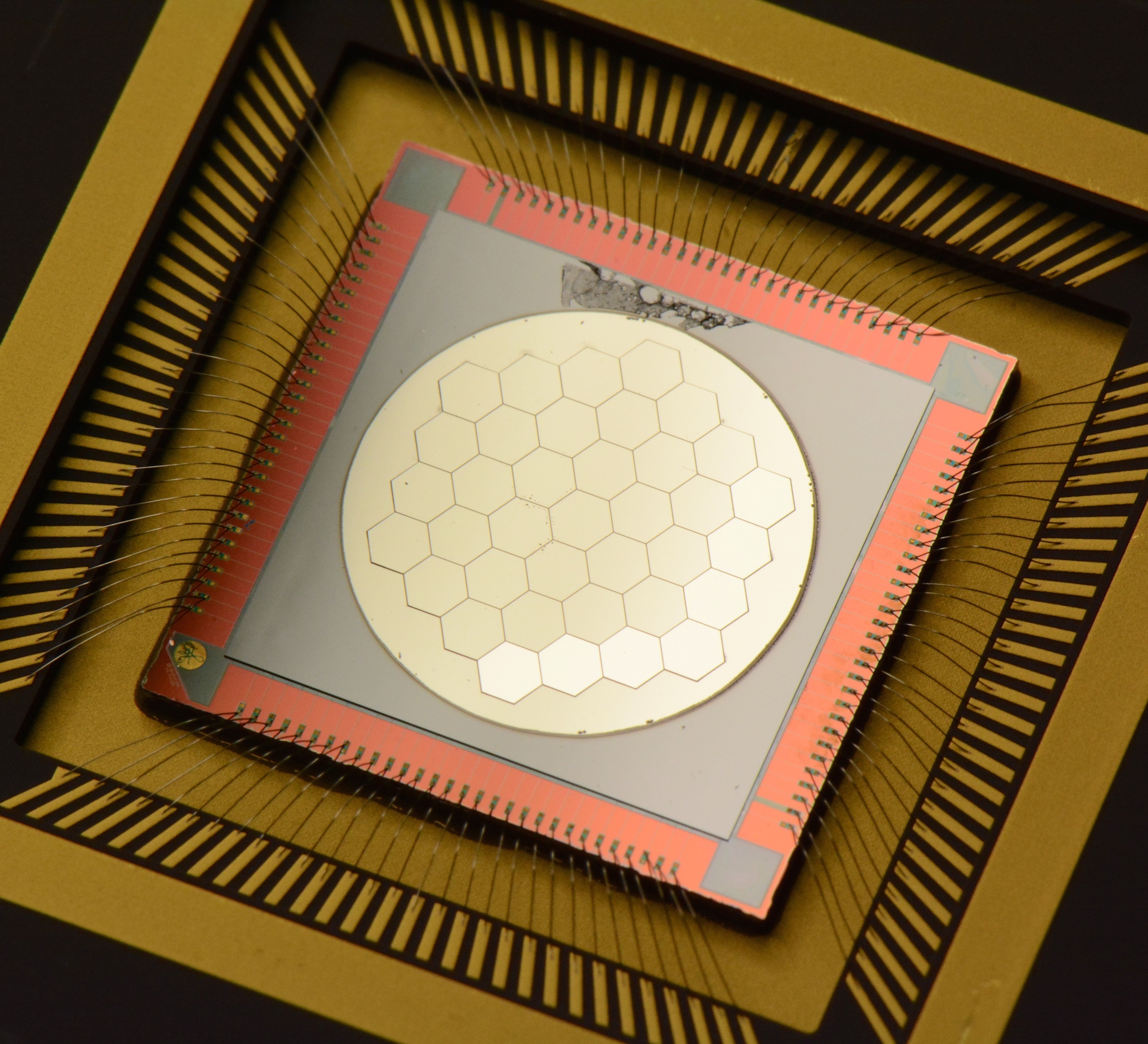}
   \caption{\label{fig:detailhardware} Photos of JOST optical elements for the surrogate JWST primary. \textit{Left:} JWST-like laser cut pupil mask. The secondary supports are about 200 microns wide. \textit{Right:} 37 segment Iris AO deformable mirror. Each segment is only 1.4 mm in size; the entire 37-segment DM is less than 1 cm across. The inter-segment gaps are 10 microns wide. Together with the L1 lens these elements implement the surrogate segmented primary mirror for our testbed.}
\end{figure}

The custom aspheric lenses for the final design were procured by late 2013 and assembled onto the optical table to within mechanical tolerances. Laser cut pupil masks were procured in 2014 to limit the aperture to 18 active segments (out of 37 on the DM) and define the central obstruction and spiders.  The optical and opto-mechanical assembly of the testbed was completed by spring 2014; see Figures \ref{fig:hardware_final} and \ref{fig:detailhardware}.  The final high precision optical alignment and software development continued into summer 2014, as discussed in the next section.

A software interface to control the testbed motors and camera was developed in Labview\footnote{http://www.ni.com/labview/}, re-using extensively the code and concepts developed for the Gemini Planet Imager Coronagraph Testbed\cite{2009SPIE.7440E..23S}. The interface also includes an autofocusing mode for the camera, which is needed as part of the alignment procedure, and will later incorporate calls to the phase retrieval and wavefront control algorithms. 

\subsection{Testbed Alignment}

A initial tolerancing analysis (see Paper II) indicated that our wavefront error requirement can be met if all optics are aligned to within $\pm$50 microns for $x,y,z$ decenters and $\pm 1.2$~ arcmin for tip/tilts.  Note that while the entire point of this testbed is to eventually be misaligning and realigning it under computer control through wavefront sensing, we wish to \textit{begin} with an optical system that has been aligned using classical techniques. This includes in particular the degrees of freedom that are not under computer control such as the position of the L1 and L3 mirrors, etc. In some ways this stage is loosely analogous to JWST being initially aligned interferometrically using the center of curvature assembly, at a stage of integration and testing prior to active wavefront control using image plane data. 

\subsubsection{Preliminary alignment with alignment telescope}
Alignment of the three lenses along a single optical axis was achieved using a Davidson Optronics D-275 autocollimating alignment telescope. This left the optics aligned in $x$ and $y$ positions, tip, and tilt, but did not constrain the $z$ positions along the optical axis since this degree of freedom cannot be sensed with this instrument. We estimate that the accuracy of the final alignment state in $x,y$, tip and tilt for all three lenses is better than 10~$\mu$m, based on the calibration and capabilities of the Davidson telescope. This is significantly better than the requirement for these degrees of freedom. We iterated the alignment procedure at both ends of the $z$ translation stage (along the optical axis) for each lens, to ensure the mechanical and optical $z$ axes are aligned such that each optic can be translated along the optical axis without adding any drift in $x,y$, tip or tilt to within a few microns (i.e. an order of magnitude below our alignment tolerance). The alignment telescope was also used to align the center of the pupil aperture onto the same optical axis.

\subsubsection{Final alignment along the optical axis}\label{coarsealgin}

\begin{figure}
 \begin{center}
 \includegraphics[trim= 0cm 6cm 0cm 6cm, clip=true, width=0.8\linewidth]{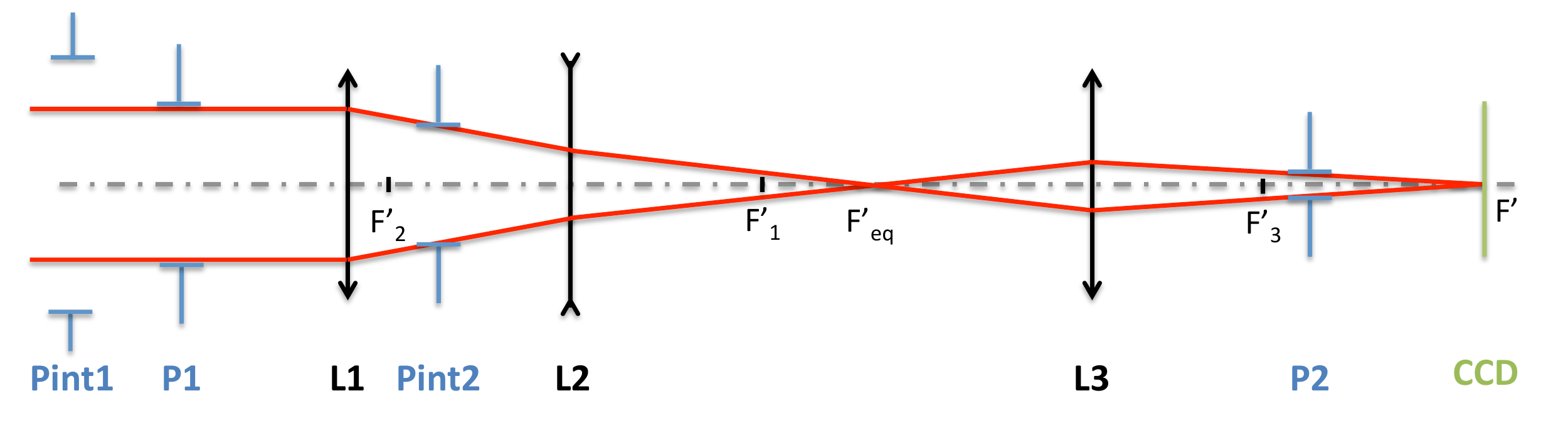}
 \end{center}
 \caption 
 {Optical scheme for the fine alignment representing the pupil P1, the lenses L1, L2 and L3, the CCD focal plane $F'$. The intermediate focal planes are represented: F$'_1$ is the image focal plane of the lens L1, F$'_2$ is the image focal plane of the lens L2, F$'_{eq}$ is the image focal plane of the system \{L1+L2\} and F$'_3$ is the image focal plane of the lens L3. The intermediate virtual images of the pupil by L1, \{L1+L2\} and \{L1+L2+L3\} are respectively represented by Pint1, Pint2 (both virtual) and P2 (real). \label{fig:alignment_procedure} }
 \end{figure}

A first coarse alignment along the optical axis is achieved by measuring and adjusting the mechanical distances between each optic and verifying the nominal geometric design parameters, which are readily derived from geometrical optic computations (see Figure \ref{fig:alignment_procedure}):
\begin{itemize}[itemsep=2pt]
\item f-ratio in the intermediate image focal plane of the system \{L1+L2\}, F$'_{eq}$
\item location and magnification of the pupil in the DM plane, P2,
\item f-ratio in the CCD image focal plane, F$'$.
\end{itemize}

A sensitivity analysis (cf. Table~\ref{table:sensitivity_analysis}) was performed with Zemax to determine the tolerance on the $z$ positions along the optical axis for each optic, in order to reach the system's WFE requirements. 

Because a global translation of the three-lens system has no effect on the WFE, we let the $z$ position of the entrance pupil define our reference plane P1, and are therefore left with three remaining unknown positions along the optical axis ($z_1$, $z_2$, $z_3$ for the three lenses). The $z$ position of the camera is also unknown but can be adjusted automatically using an auto-focus software taking advantage of the camera's motorized stage. Table~(\ref{table:sensitivity_analysis}) shows that only the distance L1-L2 has a tight $z$-positioning tolerance of 0.40~mm (WFE goal) to 0.80~mm (WFE requirement). The tolerance on the distance P1-L1 and L2-L3 are very loose, respectively to 20.2~mm and 12.2~mm (WFE goal). Therefore P1 and L1 positions can be set within their tolerances from the simple initial coarse alignment. L3 can be adjusted to deliver the correct pupil magnification at the DM plane, P2, using geometrical analysis. A translation of L2 within its WFE tolerance (0.4 to 0.8~mm) implies a translation of L3 ($<1$~mm) to recover the correct pupil magnification that is negligible compared to L3's tolerance on WFE. Therefore, we can first align L2 independently to optimize the WF over the field of view at the detector plane (refocusing the camera for each L2 move). 
Once satisfactory WFE has been obtained, a final alignment of L3 can be performed to adjust the correct pupil magnification. 

\begin{table}
 \begin{center}
 \begin{tabular}{c|rrrrr}
  & Minimum Z-axis shift & Maximum Z-axis shift & Corresponding & Maximum\\
  & from nominal position & from nominal position & Range & Wavefront Error\\
  & (mm) & (mm) & (mm) & (nm rms)\\
  \hline
  Pupil & -10.00 & 10.00 & 20.0 & 15.0 \\
  L1    &  -0.18 &  0.22 &  0.4 & 20.3 \\
  L1    &  -0.10 &  0.10 &  0.2 & 16.5 \\
  L2    &  -0.22 &  0.18 &  0.4 & 20.3 \\
  L2    &  -0.10 &  0.10 &  0.2 & 16.4 \\
  L3    & -11.00 &  4.50 & 12.0 & 15.0 \\
 \end{tabular}
 \end{center}
 \caption 
 {Sensitivity analysis for the Z-axis positions of the pupil, L1, L2 and L3. "Nominal position" refers to the corresponding theoretical position given by the Zemax software.\label{table:sensitivity_analysis} }
 \end{table}

\subsection{First results}

\begin{figure}
 \centering
  \includegraphics[height=0.45\textwidth]{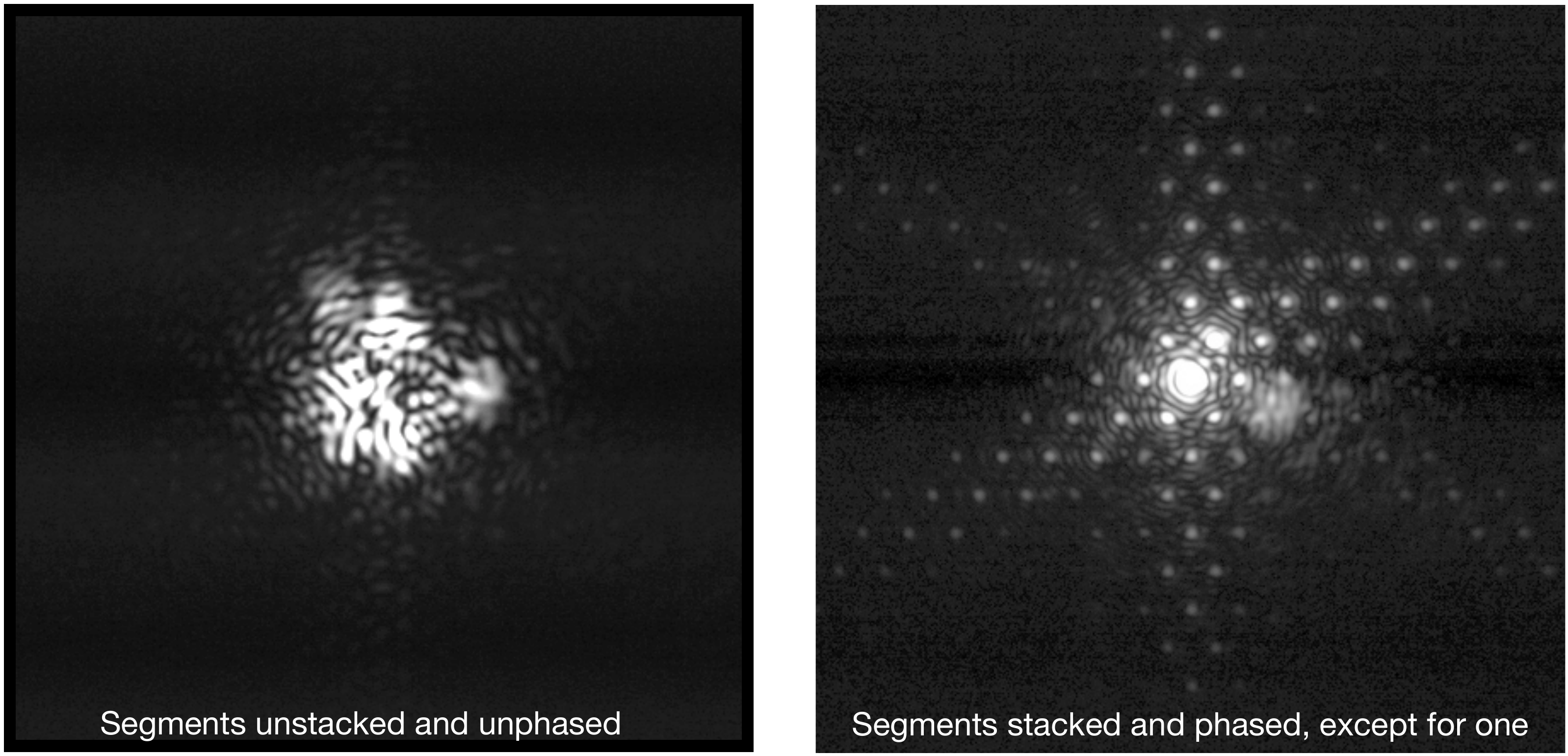}
 \caption{\label{fig:stacking} Example of point spread functions observed with JOST and the Iris AO segmented deformable mirror, from the initial preliminary testbed phase. \textit{Left:} Point spread function with all segments unstacked and unphased. \textit{Right:} PSF with almost all segments stacked and cophased. These data were taken using the initial engineering-grade DM with one nonfunctional segment, which is responsible for the spot of extra off-axis light to the lower right of the PSF center.}
\end{figure}

\begin{figure}
 \centering
  \includegraphics[trim= 2cm 5.8cm 2cm 5.8cm, clip=true, height=0.4\textwidth]{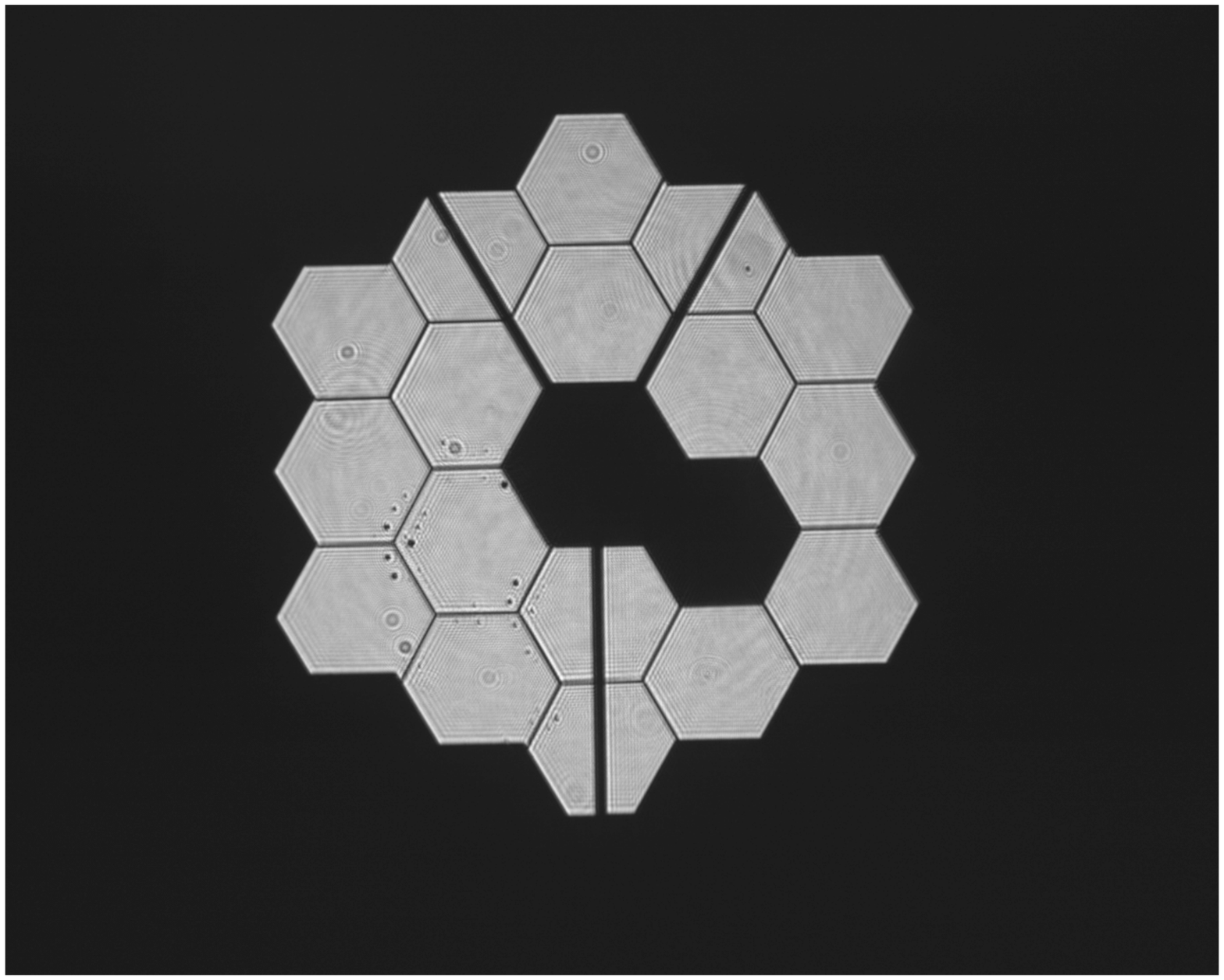}
  \includegraphics[height=0.4\textwidth]{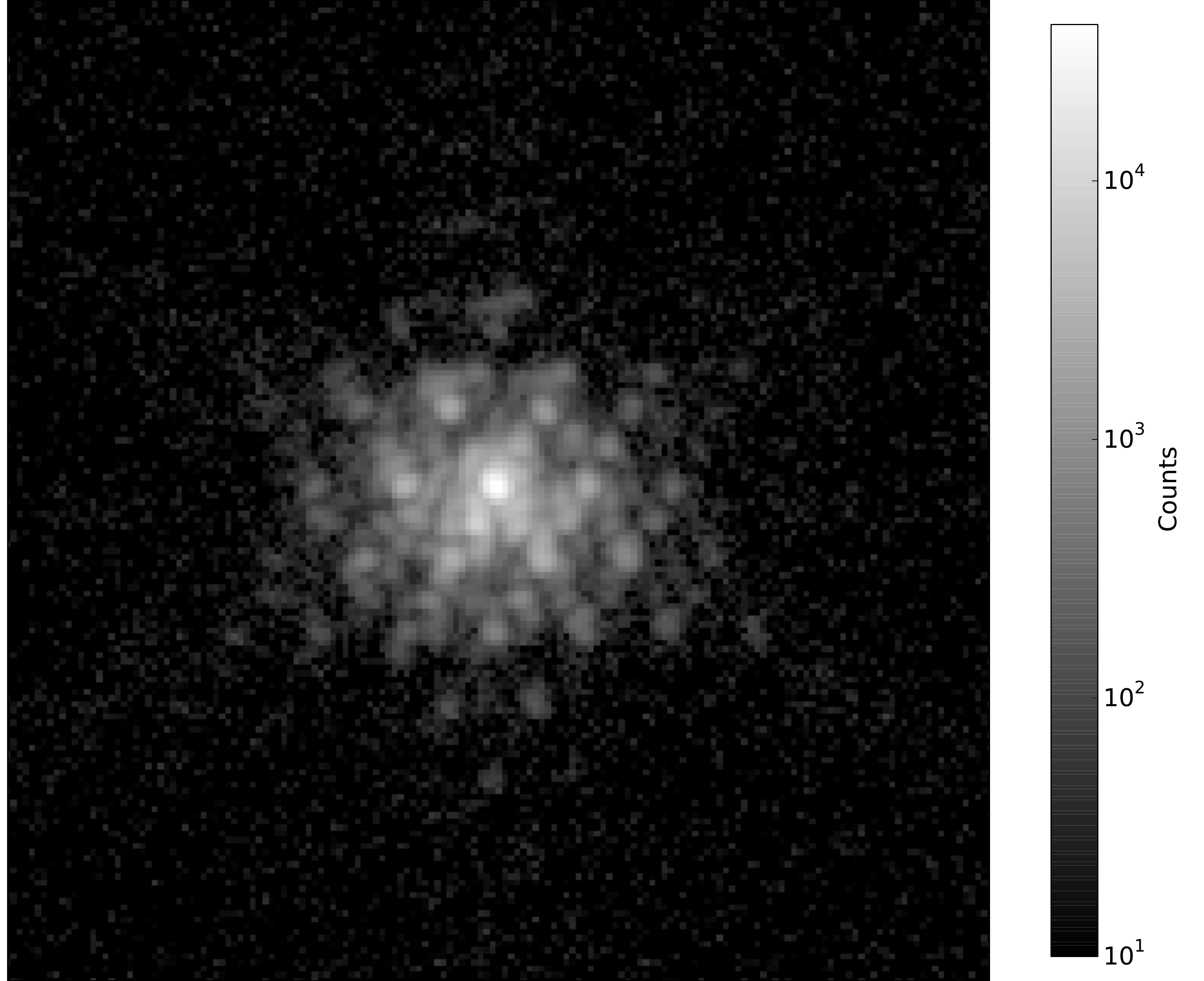}
 \caption{\label{fig:pupilimg}\textit{Left:} Pupil viewer image of the IRIS AO segmented deformable mirror with the JWST mask. The additional blank segment adjacent to the center is due to a nonfunctional segment in the current engineering-grade DM. A science grade DM with fully functional actuators will be available this fall. The particular pupil mask shown here is slightly undersized to block pupil wander, so the illuminated regions of the outer segments are by design no longer precisely hexagonal. \textit{Right:} PSF of the segmented deformable mirror in our lab. The segments are approximately stacked but not yet perfectly phased in this image.}
\end{figure}

Using the preliminary testbed implementation and the engineering IRIS-AO DM, we first demonstrated the phasing capabilities in open loop using the factory calibration (Figure~\ref{fig:stacking}) using all 37 DM segments (i.e. operating without the JWST pupil).  More recently, we have implemented the JWST pupil masks and obtained first pupil images and single field point focal images through the complete and coarsely aligned system, as shown in Figure~\ref{fig:pupilimg}. Note the effect of the uncontrollable engineering-grade segment in the pupil image. The final science grade DM will be implemented in the fall 2014. The quality of the PSF is commensurate with the expectations for this phase of the alignment. 

\begin{figure}
 \centering
  \includegraphics[height=0.18\textwidth]{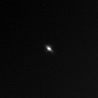}
  \includegraphics[height=0.18\textwidth]{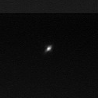}
  \includegraphics[height=0.18\textwidth]{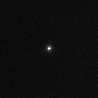}
  \includegraphics[height=0.18\textwidth]{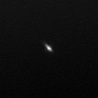}
  \includegraphics[height=0.18\textwidth]{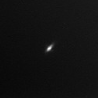}
 \caption{\label{fig:fovresults} Set of 5 images (center + corners) after coarse alignment but before fine alignment of the final testbed version using the custom aspheric optics. The on-axis central PSF is diffraction limited, but the corners of the field of view show radially elongated PSFs due to $z$ axis misalignment of the powered optics. Completion of the fine optical alignment to achieve diffraction limited PSF quality over the entire field of view is expected in the immediate future as of this writing. }
\end{figure}

In Figure \ref{fig:fovresults} we show the PSF at the center and four corners of the field of view, after focusing the detector at the center of the field. This set of images is obtained after the coarse alignment phase described in Section \ref{coarsealgin}. Visual analysis of these PSF show that they are dominated by astigmatism and field curvature given the symmetry of the PSF images, as we can expect from a coarse alignment along the z-axis. This observed symmetry also confirms at this coarse stage that the alignment of $x,y$ translation modes and tip/tilt modes was successful. 

\section{CONCLUSION, AND FUTURE ACTIVITIES}

We have presented an overview of the JWST Optical Simulator Testbed including its main goals, requirements, and first results. Designed as a versatile WFS\&C experiment platform, JOST will complement simulation studies (e.g.\ using BATC's high-fidelity JWST-ITM package \cite{Knight2012SPIE.8449E..0VK}), by enabling independent experiments that cross-check models with actual data from a simplified and compact testbed that nonetheless reproduces the key optical physics relevant for JWST WFS\&C.

In particular, the 59 controllable optical degrees of freedom for JOST---piston, tip, and tilt for the primary segments and piston, tip, tilt, and $x$- and $y$-decenter for the secondary---are precisely the degrees of freedom for JWST that are expected to be controlled periodically during routine maintenance of the observatory's alignment. The segmented mirror degrees of freedom not controllable for JOST---radial and azimuthal decentering, clocking, and radius of curvature---are those with the smallest influence functions in JWST; these are expected to need adjustment very rarely, if ever, after the completion of the Global Alignment stage of JWST commissioning. Furthermore, JOST cannot replicate all the sensing modalities used in JWST commissioning, for instance neither coarse phasing using a dispersed Hartmann sensor (DHS) nor MIMF sensing using broadband PSFs for e.g. the FGS, but it does implement carefully the sensing mode that will be used for routine maintenance: focus diverse phase retrieval using NIRCam at one or more field points. JOST thus provides a highly capable platform for modeling the routine maintenance of JWST's alignment over the long term, precisely the task for which the Telescopes Team at STScI will have primary responsibility. For OTE commissioning, the primary responsibility is with our colleagues at Ball Aerospace and Northrup Grumman, and commissioning methods have been verified using the TBT and ITM and will be further validated during upcoming ISIM and OTIS test campaigns. JOST nonetheless is also capable of a range of investigations related to commissioning activities, without having any necessity for it to be able to simulate every aspect of commissioning JWST.


As such, JOST is not part of the main JWST development path but rather a tool for independent verification and cross checks, and experiments with new techniques. We will use the testbed mostly for independent studies and validations of key aspects of the JWST WFS\&C flow, and for developing contingency plans or enhancements to the baseline via new algorithms (e.g. improved phase retrieval). 
In some cases it may be possible to use JOST to generate data to assist with or accelerate integration and test activities for portions of the ground system. We will evaluate on a case-by-case basis which tasks are better suited to pure simulation versus making use of the testbed. These goals and activities will contribute to the oversight role of the Telescope Scientist (M. Mountain). 

The long term testbed goals are organized in several phases and themes:

\textbf{i) Initial testbed assembly and alignment:}
This phase is being completed at the time of this publication, as detailed above. 
A precisely aligned state of the system is expected to be reached in the summer of 2014. 

\textbf{ii) Wavefront sensing experiments:}
We will proceed to the implementation of phase retrieval code, and its integration and automation in the JOST Labview interface. The first step will be to implement single field point phase retrieval, using the open-loop factory calibration of the Iris AO DM to introduce various wavefront states. In particular we will implement the Hybrid Diversity Algorithm\cite{Dean2006HDA} developed for  JWST fine wavefront sensing phase retrieval. Our 4D Fizeau interferometer will be used for direct measurement of the wavefront and to validate the phase retrieval algorithm performance. Multi-field point phase retrieval will then be used to verify and refine the alignment state of the testbed. 

JOST will provide a flexible testbed for investigating different forms of phase retrieval, including recent developments in parametric phase retrieval and non-linear optimization\cite{Fienup:93,Brady:06}, transverse translational diversity\cite{Brady:09} and related approaches \cite{2012SPIE.8447E..6PC}. 

\textbf{iii) Single field point wavefront control: }
The next step in fidelity is to achieve active control of the segmented DM and position of the secondary based on the phase retrieval results. WFS\&C will be implemented using a linear model of the system, which must be calculated specifically for our testbed since its control matrices will differ in detail from the JWST ones; the reduced degrees of freedom and flat segments result in this being a more straightforward problem than the corresponding linear control model for JWST itself.  We will investigate the performance of the system using control matrices from an optical ray tracing model of the testbed\cite{Choquet2014}, and by building empirical control matrices from testbed data. Using software to intentionally inject changing mirror states as a function of time, we can simulate active control of a dynamic time-variable space observatory. 

\textbf{iv) Commissioning studies:}
The testbed will be adjusted to an estimated post-­deployment state by moving optics and DM actuators. The goal will be to re­align the testbed following the commissioning sequence, or at least its main phases. The most interesting aspects of this goals is the implementation of multiple field point WFS\&C, over a field of view equivalent to that of NIRCam. 
Typical experiments will involve introducing misalignments into the system (both on the DM and on the secondary mirror surrogate) followed by sensing and control activities. Other experiments will be designed to reproduce certain phases of the JWST commissioning compatible with the available degrees of freedom and travel ranges on JOST. While it will not be possible to replicate with JOST the full diversity of optical states and misalignments that will be encountered in the course of JWST commissioning, a sufficient number of interesting scenarios are possible.

\textbf{v) Integration and Test (I\&T) activities:} We also plan to make some limited use of the actual JWST WFS\&C Analysis Software\cite{Acton2012SPIE_JWSTWFSC_Overview} (WAS) for wavefront sensing analysis of JOST data. The testbed was designed to provide the exact same sampling as NIRCam at the WFS wavelength to enable this capability. A software interface will be developed to format JOST testbed data in such a fashion that it can be used directly as input to the WAS, aided by some adjustment of WAS configuration file parameters as necessary. We will compare results between the WAS results, our own implementation of the HDA algorithm, and with the direct Fizeau measurement.

Once JOST data can be interfaced directly with ground systems at the S\&OC, we can potentially use the testbed for various I\&T activites, including additional testing of the WFS\&C software subsystem (WSS) based on real data, and even potentially passing data through other portions of the SOC ground system to verify data flow.  One potential advantage of JOST is that data can be obtained quasi-instantly by contrast to high-fidelity software simulations (e.g. ITM) that require more significant computing time to generate simulated data. 

\textbf{v) Contingency planning, and other studies:}
An additional goal will be to build confidence in various contingency cases and backup plans for JWST, for instance tests to validate the proposed non­redundant tilts backup mirror phasing strategy\cite{2012OExpr..2029457C} or implementation of recent phase retrieval techniques with extended capture range\cite{2014JOSAA..31..661J}. Much remains to be done in developing these plans for the highest priority cases under consideration. JOST's flexibility as a general-purpose wavefront sensing testbed will provide a platform for addressing such questions as needed.

JWST's prime mission lifetime is 5 years after commissioning, with a goal of 10 years. The state-of-the-art in wavefront sensing techniques will assuredly continue to evolve during that time.  Furthermore, it is reasonably likely that over time, new staff members will rotate onto the telescope optics team responsible for JWST's wavefront maintenance. JOST now provides an in-house facility for developing staff expertise in wavefront sensing and control techniques at STScI to ensure JWST's active optics will be expertly looked after and optimized for the lifetime of the observatory.


\acknowledgments   
 
M.D.P. and R.S. conceived of this project and led the overall effort. O.L., E.C. and M.N. developed and implemented the optical and mechanical design. O.L., E.C., M.N. and C.-P.L. assembled the testbed and performed preliminary alignment, while M.Y., L.L. and S.E. performed the fine alignment.  C.-P.L., O.L. and M.Y. developed the control software. C.L. helped with assembly and machining. As lab manager R.A. coordinated support and lab infrastructure for the new testbed.   E.E., G.H., L.P., and R.vdM. provided valuable advice and guidance. M.M. enabled this project through his JWST Telescope Scientist grant and support of the creation of the STScI Russell B. Makidon Optics Laboratory.  

The work reported here was supported by NASA grant NNX07AR82G (PI: Mountain). 
We thank Michael Helmbrecht at Iris AO for searching for an engineering grade DM with a ``JWST-like'' clear region of 18 good segments.
We thank Scott Acton, Scott Knight, Bruce Dean, and Tom Zielinski for interesting discussions on the design and on wavefront sensing methods more generally. We are grateful to Kelly Coleman and Robin Auer for assistance in procurement of the custom optics, DMs, and other laboratory equipment, and to Joe Hunkeler for assistance with software development for controlling the SBIG CCD.
We thank Michael McElwain and Qian Gong at NASA GSFC for the loan of the alignment telescope and helpful discussions.


\bibliography{report}  

\begin{thebibliography}{10}

\bibitem{Acton2004SPIE.5487..887A}
{Acton}, D.~S., {Atcheson}, P.~D., {Cermak}, M., {Kingsbury}, L.~K., {Shi}, F.,
  and {Redding}, D.~C., ``{James Webb Space Telescope wavefront sensing and
  control algorithms},'' in [{\em Optical, Infrared, and Millimeter Space
  Telescopes}{\nolinebreak\hspace{0.1em}]},  {Mather}, J.~C., ed., {\em
  \procspie} {\bf 5487},  887--896 (Oct. 2004).

\bibitem{Acton2012SPIE_JWSTWFSC_Overview}
{Acton}, D.~S., {Knight}, J.~S., {Contos}, A., {Grimaldi}, S., {Terry}, J.,
  {Lightsey}, P., {Barto}, A., {League}, B., {Dean}, B., {Smith}, J.~S.,
  {Bowers}, C., {Aronstein}, D., {Feinberg}, L., {Hayden}, W., {Comeau}, T.,
  {Soummer}, R., {Elliott}, E., {Perrin}, M., and {Starr}, C.~W., ``{Wavefront
  sensing and controls for the James Webb Space Telescope},'' in [{\em
  \procspie}{\nolinebreak\hspace{0.1em}]},  {\em \procspie} {\bf 8442} (Sept.
  2012).

\bibitem{Knight2012SPIE.8442E..2CK}
{Knight}, J.~S., {Acton}, D.~S., {Lightsey}, P., {Contos}, A., and {Barto}, A.,
  ``{Observatory alignment of the James Webb Space Telescope},'' in [{\em
  \procspie}{\nolinebreak\hspace{0.1em}]},  {\em \procspie} {\bf 8442} (Sept.
  2012).

\bibitem{Spergel2013arXiv1305.5422S}
{Spergel}, D., {Gehrels}, N., {Breckinridge}, J., {Donahue}, M., {Dressler},
  A., {Gaudi}, B.~S., {Greene}, T., {Guyon}, O., {Hirata}, C., {Kalirai}, J.,
  {Kasdin}, N.~J., {Moos}, W., {Perlmutter}, S., {Postman}, M., {Rauscher}, B.,
  {Rhodes}, J., {Wang}, Y., {Weinberg}, D., {Centrella}, J., {Traub}, W.,
  {Baltay}, C., {Colbert}, J., {Bennett}, D., {Kiessling}, A., {Macintosh}, B.,
  {Merten}, J., {Mortonson}, M., {Penny}, M., {Rozo}, E., {Savransky}, D.,
  {Stapelfeldt}, K., {Zu}, Y., {Baker}, C., {Cheng}, E., {Content}, D.,
  {Dooley}, J., {Foote}, M., {Goullioud}, R., {Grady}, K., {Jackson}, C.,
  {Kruk}, J., {Levine}, M., {Melton}, M., {Peddie}, C., {Ruffa}, J., and
  {Shaklan}, S., ``{Wide-Field InfraRed Survey Telescope-Astrophysics Focused
  Telescope Assets WFIRST-AFTA Final Report},'' {\em ArXiv e-prints}  (May
  2013).

\bibitem{Postman2009arXiv0904.0941P}
{Postman}, M., {Argabright}, V., {Arnold}, B., {Aronstein}, D., {Atcheson}, P.,
  {Blouke}, M., {Brown}, T., {Calzetti}, D., {Cash}, W., {Clampin}, M.,
  {Content}, D., {Dailey}, D., {Danner}, R., {Doxsey}, R., {Ebbets}, D.,
  {Eisenhardt}, P., {Feinberg}, L., {Fruchter}, A., {Giavalisco}, M.,
  {Glassman}, T., {Gong}, Q., {Green}, J., {Grunsfeld}, J., {Gull}, T.,
  {Hickey}, G., {Hopkins}, R., {Hraba}, J., {Hyde}, T., {Jordan}, I., {Kasdin},
  J., {Kendrick}, S., {Kilston}, S., {Koekemoer}, A., {Korechoff}, B., {Krist},
  J., {Mather}, J., {Lillie}, C., {Lo}, A., {Lyon}, R., {McCullough}, P.,
  {Mosier}, G., {Mountain}, M., {Oegerle}, B., {Pasquale}, B., {Purves}, L.,
  {Penera}, C., {Polidan}, R., {Redding}, D., {Sahu}, K., {Saif}, B.,
  {Sembach}, K., {Shull}, M., {Smith}, S., {Sonneborn}, G., {Spergel}, D.,
  {Stahl}, P., {Stapelfeldt}, K., {Thronson}, H., {Thronton}, G., {Townsend},
  J., {Traub}, W., {Unwin}, S., {Valenti}, J., {Vanderbei}, R., {Werner}, M.,
  {Wesenberg}, R., {Wiseman}, J., and {Woodgate}, B., ``{Advanced Technology
  Large-Aperture Space Telescope (ATLAST): A Technology Roadmap for the Next
  Decade},'' {\em ArXiv e-prints}  (Apr. 2009).

\bibitem{Barto2008SPIE.7010E..23B}
{Barto}, A.~A., {Atkinson}, C., {Contreras}, J., {Lightsey}, P.~A., {Noecker},
  C., {Waldman}, M., and {Whitman}, T., ``{Optical performance verification of
  the James Webb Space Telescope},'' in [{\em
  \procspie}{\nolinebreak\hspace{0.1em}]},  {\em \procspie} {\bf 7010} (Aug.
  2008).

\bibitem{Acton2006SPIE.6265E..21A}
{Acton}, D.~S., {Towell}, T., {Schwenker}, J., {Swensen}, J., {Shields}, D.,
  {Sabatke}, E., {Klingemann}, L., {Contos}, A.~R., {Bauer}, B., {Hansen}, K.,
  {Atcheson}, P.~D., {Redding}, D., {Shi}, F., {Basinger}, S., {Dean}, B., and
  {Burns}, L., ``{Demonstration of the James Webb Space Telescope commissioning
  on the JWST testbed telescope},'' in [{\em
  \procspie}{\nolinebreak\hspace{0.1em}]},  {\em \procspie} {\bf 6265} (July
  2006).

\bibitem{Acton2007SPIE.6687E...5A}
{Acton}, D.~S., {Towell}, T., {Schwenker}, J., {Shields}, D., {Sabatke}, E.,
  {Contos}, A.~R., {Hansen}, K., {Shi}, F., {Dean}, B., and {Smith}, S.,
  ``{End-to-end commissioning demonstration of the James Webb Space
  Telescope},'' in [{\em \procspie}{\nolinebreak\hspace{0.1em}]},  {\em
  \procspie} {\bf 6687} (Sept. 2007).

\bibitem{Knight2012SPIE.8449E..0VK}
{Knight}, J.~S., {Acton}, D.~S., {Lightsey}, P., and {Barto}, A., ``{Integrated
  telescope model for the James Webb Space Telescope},'' in [{\em
  \procspie}{\nolinebreak\hspace{0.1em}]},  {\em \procspie} {\bf 8449} (Sept.
  2012).

\bibitem{Sivaramakrishnan2012SPIE.8442E..2SS}
{Sivaramakrishnan}, A., {Lafreni{\`e}re}, D., {Ford}, K.~E.~S., {McKernan}, B.,
  {Cheetham}, A., {Greenbaum}, A.~Z., {Tuthill}, P.~G., {Lloyd}, J.~P.,
  {Ireland}, M.~J., {Doyon}, R., {Beaulieu}, M., {Martel}, A., {Koekemoer}, A.,
  {Martinache}, F., and {Teuben}, P., ``{Non-redundant Aperture Masking
  Interferometry (AMI) and segment phasing with JWST-NIRISS},'' in [{\em
  \procspie}{\nolinebreak\hspace{0.1em}]},  {\em \procspie} {\bf 8442} (Sept.
  2012).

\bibitem{Jurling2012SPIE.8442E..10J}
{Jurling}, A.~S. and {Content}, D.~A., ``{Wavefront sensing for WFIRST with a
  linear optical model},'' in [{\em \procspie}{\nolinebreak\hspace{0.1em}]},
  {\em \procspie} {\bf 8442} (Sept. 2012).

\bibitem{Pope2014MNRAS.440..125P}
{Pope}, B., {Cvetojevic}, N., {Cheetham}, A., {Martinache}, F., {Norris}, B.,
  and {Tuthill}, P., ``{A demonstration of wavefront sensing and mirror phasing
  from the image domain},'' {\em \mnras}~{\bf 440},  125--133 (May 2014).

\bibitem{Jurling:2014}
Jurling, A.~S. and Fienup, J.~R., ``Extended capture range for focus-diverse
  phase retrieval in segmented aperture systems using geometrical optics,''
  {\em \josaa}~{\bf 31},  661 (Mar. 2014).

\bibitem{Choquet2014}
{Choquet}, E., {Levecq}, O., {N'Diaye}, M., {Perrin}, M.~D., and {Soummer}, R.,
  ``{James Webb Space Telescope Optical Simulation Testbed II. Design of a
  Three-Lens Anastigmat Telescope Simulator},'' in [{\em
  \procspie}{\nolinebreak\hspace{0.1em}]},  {\em \procspie} {\bf 9143} (2014).

\bibitem{Acton2012SPIE_MIMF}
{Acton}, D.~S. and {Knight}, J.~S., ``{Multi-field alignment of the James Webb
  Space Telescope},'' in [{\em \procspie}{\nolinebreak\hspace{0.1em}]},  {\em
  \procspie} {\bf 8442} (Sept. 2012).

\bibitem{2013SPIE.8864E..1KN}
{N'Diaye}, M., {Choquet}, E., {Pueyo}, L., {Elliot}, E., {Perrin}, M.~D.,
  {Wallace}, J.~K., {Groff}, T., {Carlotti}, A., {Mawet}, D., {Sheckells}, M.,
  {Shaklan}, S., {Macintosh}, B., {Kasdin}, N.~J., and {Soummer}, R.,
  ``{High-contrast imager for complex aperture telescopes (HiCAT): 1. testbed
  design},'' in [{\em \procspie}{\nolinebreak\hspace{0.1em}]},  {\em \procspie}
  {\bf 8864} (Sept. 2013).

\bibitem{2009SPIE.7440E..23S}
{Soummer}, R., {Sivaramakrishnan}, A., {Oppenheimer}, B.~R., {Roberts}, R.,
  {Brenner}, D., {Carlotti}, A., {Pueyo}, L., {Macintosh}, B., {Bauman}, B.,
  {Saddlemyer}, L., {Palmer}, D., {Erickson}, D., {Dorrer}, C., {Caputa}, K.,
  {Marois}, C., {Wallace}, K., {Griffiths}, E., and {Mey}, J., ``{The Gemini
  Planet Imager coronagraph testbed},'' in [{\em
  \procspie}{\nolinebreak\hspace{0.1em}]},  {\em \procspie} {\bf 7440} (Aug.
  2009).

\bibitem{Dean2006HDA}
Dean, B.~H., Aronstein, D.~L., Smith, J.~S., Shiri, R., and Acton, D.~S.,
  ``Phase retrieval algorithm for jwst flight and testbed telescope,'' {\em
  Proc. SPIE}~{\bf 6265},  626511--626511--17 (2006).

\bibitem{Fienup:93}
Fienup, J.~R., ``Phase-retrieval algorithms for a complicated optical system,''
  {\em Appl. Opt.}~{\bf 32},  1737--1746 (Apr 1993).

\bibitem{Brady:06}
Brady, G.~R. and Fienup, J.~R., ``Nonlinear optimization algorithm for
  retrieving the full complex pupil function,'' {\em Opt. Express}~{\bf 14},
  474--486 (Jan 2006).

\bibitem{Brady:09}
Brady, G.~R., Guizar-Sicairos, M., and Fienup, J.~R., ``Optical wavefront
  measurement using phase retrieval with transverse translation diversity,''
  {\em Opt. Express}~{\bf 17},  624--639 (Jan 2009).

\bibitem{2012SPIE.8447E..6PC}
{Codona}, J.~L., ``{Theory and application of differential OTF (dOTF) wavefront
  sensing},'' in [{\em Society of Photo-Optical Instrumentation Engineers
  (SPIE) Conference Series}{\nolinebreak\hspace{0.1em}]},  {\em Society of
  Photo-Optical Instrumentation Engineers (SPIE) Conference Series} {\bf 8447}
  (July 2012).

\bibitem{2012OExpr..2029457C}
{Cheetham}, A.~C., {Tuthill}, P.~G., {Sivaramakrishnan}, A., and {Lloyd},
  J.~P., ``{Fizeau interferometric cophasing of segmented mirrors},'' {\em
  Optics Express}~{\bf 20},  29457 (Dec. 2012).

\bibitem{2014JOSAA..31..661J}
{Jurling}, A.~S. and {Fienup}, J.~R., ``{Extended capture range for
  focus-diverse phase retrieval in segmented aperture systems using geometrical
  optics},'' {\em Journal of the Optical Society of America A}~{\bf 31},  661
  (Mar. 2014).

\end{thebibliography}
\bibliographystyle{spiebib}  

\end{document}